\documentstyle[12pt]{article}
\begin{document} 
\global\parskip 6pt
\newcommand{\be}{\begin{equation}}
\newcommand{\ee}{\end{equation}}
\newcommand{\bea}{\begin{eqnarray}}
\newcommand{\eea}{\end{eqnarray}}
\newcommand{\non}{\nonumber}

\begin{titlepage}
\hfill{hep-th/9808032}
\vspace*{1cm}
\begin{center}
{\Large\bf Topological Black Holes in Anti-de Sitter Space}\\
\vspace*{2cm}
Danny Birmingham\footnote{E-mail: dannyb@pop3.ucd.ie}\\
\vspace*{.5cm}
{\em Department of Mathematical Physics\\
University College Dublin\\
Belfield, Dublin 4, Ireland}\\
\vspace{2cm}

\begin{abstract}
We consider a class of black hole solutions to Einstein's equations
in $d$ dimensions with a negative cosmological constant.
These solutions have the property that
the horizon is a $(d-2)$-dimensional Einstein manifold of
positive, zero, or negative curvature.
The mass, temperature, and entropy are calculated.
Using the correspondence with conformal field theory, the phase
structure of the
solutions is examined, and used to determine the correct
mass dependence of the Bekenstein-Hawking entropy.
\end{abstract}
\vspace{1cm}
August 1998 \\
\end{center}
\end{titlepage}

\section{Introduction}
Recently, a great deal of attention has been paid to a conjectured
correspondence between string theory (and supergravity) defined
on direct products of $d$-dimensional anti-de Sitter space
with compact manifolds,
and the large $N$ limit of certain conformal field theories
defined on the $(d-1)$-dimensional boundary of the anti-de Sitter
space \cite{Malda}. This followed earlier
investigations of scattering from branes \cite{Gub}-\cite{Strom},
and near-horizon brane geometry
\cite{Gibbons,Duff}.
The precise nature of this correspondence was explored in
\cite{Poly,Witten1}.

In \cite{Witten1,Witten2}, this correspondence was used to study the
boundary conformal field theory at finite temperature. In such a case,
one is studying the $SU(N)$ gauge theory at large $N$ on a manifold
$S^{1} \times M^{d-2}$.
The relevant configurations on the anti-de Sitter
side of the correspondence are then black hole solutions
to Einstein's equations with a negative cosmological
constant.
In particular, the case of $M^{d-2} = S^{d-2}$ was studied, using
the black hole solutions with spherical horizon constructed in
\cite{Carter}. The thermodynamic properties of these
solutions were discussed in
\cite{HP}.
Given the importance of this conjecture for understanding
the large $N$ dynamics of gauge theory, it is useful
to study several examples.
According to the formulation presented in \cite{Witten1},
the partition function of the conformal field theory
at finite temperature, defined on $S^{1} \times M^{d-2}$,
is given by summing
the exponential of the supergravity action
over Einstein manifolds with negative
cosmological constant which have boundary $S^{1} \times M^{d-2}$.

In this paper, we consider a class of static black hole solutions to
Einstein's equations in $d$ dimensions
with a negative cosmological constant, which have the property
that the horizon $M^{d-2}$ is a $(d-2)$-dimensional compact
Einstein space of positive, zero, or negative curvature.
These solutions are obtained by
a straightforward generalization of the four-dimensional
ansatz presented in \cite{Mann}-\cite{Vanzo}.
However, we point out that the $d$-dimensional
generalization of this ansatz only requires the horizon to be
Einstein.
Among these solutions, one has a subclass consisting of black
holes which are asymptotically locally anti-de Sitter; in this case,
the horizon has constant curvature.
We examine the horizon structure, and compute the mass, temperature,
and entropy. Within the framework of the adS/CFT correspondence,
these solutions allow us to study the conformal field theory at
finite temperature
defined on manifolds $S^{1} \times M^{d-2}$.
We investigate the phase structure of these solutions
in the light of this correspondence, and show how this leads
to a microscopic derivation of the correct mass dependence
of the Bekenstein-Hawking entropy formula.

\section{Construction of Black Hole Solutions}
Black hole solutions to Einstein's equations
with a negative cosmological constant and with spherical horizon
topology were constructed in \cite{Carter}, and their thermodynamic
properties
were were investigated in \cite{HP}.
In four dimensions, black holes for which the topology
of the horizon is an arbitrary genus Riemann surface have
been constructed in \cite{Mann}-\cite{Klemm};
the thermodynamics of these solutions has been considered
\cite{Brill2,Vanzo}.
This followed earlier work on the planar and toroidal case
\cite{Lemos1}-\cite{Cai}.
Higher-dimensional constant curvature black holes with
negative cosmological constant were
obtained in \cite{Mann2,Ban}.

Our goal here is to consider the generalization to $d$ dimensions
of the metric ansatz presented in \cite{Mann}-\cite{Vanzo}.
In order to construct solutions to Einstein's equations with
a negative cosmological constant, we begin with the metric ansatz
\bea
ds^{2} = -f(r)dt^2 + f^{-1}(r)dr^2
+ r^2 h_{ij}(x)dx^{i}dx^{j},
\label{ansatz}
\eea
where the coordinates are labelled as $x^{\mu} = (t,r,x^{i}),
(i=1,...,(d-2))$. The metric function $h_{ij}$
is a function of the coordinates $x^{i}$ only, and we shall
refer to this metric as the horizon metric. We take the horizon to be a
compact orientable manifold denoted by $M^{d-2}$.
Adopting the curvature conventions of \cite{HE},
we determine the non-vanishing components
of the Ricci tensor to be
\bea
R_{tt} &=& \frac{1}{2} f f^{\prime\prime}
+ \frac{1}{2r} (d-2)ff^{\prime},\non\\
R_{rr} &=& -\frac{1}{2} \frac{f^{\prime\prime}}{f} - \frac{1}{2r} (d-2)
\frac{f^{\prime}}{f},\non\\
R_{ij} &=& R_{ij}(h) -h_{ij}[(d-3)f + rf^{\prime}],
\label{ricci}
\eea
where $R_{ij}(h)$ is the Ricci tensor of the horizon metric,
and $f^{\prime} = df/dr$.
Let us now take the function $f$ to be of the form
\bea
f = k -\frac{\omega_{d}M}{r^{d-3}} + \frac{r^{2}}{l^{2}},
\label{f}
\eea
where
\bea
\omega_{d} = \frac{16 \pi G}{(d-2)\mathrm{Vol}(M^{d-2})},
\label{omega}
\eea
and $\mathrm{Vol}(M^{d-2}) = \int d^{d-2}x\;\sqrt{h}$.
Here, k is as yet undetermined; $l$ is a parameter with dimensions
of length, and $\omega_{d}$ is inserted for convenience so that
the parameter $M$ has dimensions of inverse length. This metric ansatz
is the $d$-dimensional generalization of that given in
\cite{Mann}-\cite{Vanzo}.
With this form of $f$, one can now check that the spacetime
is an Einstein space with negative cosmological constant, namely
\bea
R_{\mu\nu} = -\frac{(d-1)}{l^{2}}g_{\mu\nu},
\label{rmunu}
\eea
provided the horizon is an Einstein space of the form
\bea
R_{ij}(h) = (d-3)k\;h_{ij}.
\label{rij}
\eea
It is important to observe here that we have obtained a solution to
Einstein's equations with a negative cosmological constant for
any value of $k$, provided the horizon is itself Einstein.
However, the horizon may be an Einstein space
with positive, zero, or negative curvature. This opens up the possibility
to construct black hole solutions in which the topology of the horizon
is non-spherical. In particular, compact Einstein spaces of non-constant
curvature exist provided $(d-2) > 3$, see \cite{Carlip3}, for example.

However, for the moment let us continue and analyse the
structure of the Riemann tensor of these solutions.
The non-vanishing components are given by
\bea
R_{trtr} &=& \frac{1}{2}f^{\prime\prime},\;\;
R_{titj} = \frac{r}{2} ff^{\prime}\; h_{ij},\;\;
R_{rirj} = -\frac{r}{2}\frac{f^{\prime}}{f}\; h_{ij},\non\\
R_{ijkl} &=& r^{2}R_{ijkl}(h) - r^{2}f [h_{ik}\;h_{jl}
-h_{il}\;h_{jk}],
\label{riemann}
\eea
where $R_{ijkl}(h)$ is the Riemann tensor constructed from the
horizon metric $h_{ij}$.
We see that $M=0$ solution is locally isometric to anti-de Sitter
space (i.e., a spacetime of constant negative curvature),
\bea
R_{\mu\nu\rho\sigma} = -\frac{1}{l^{2}}(g_{\mu\rho}\;g_{\nu\sigma}
-g_{\mu\sigma}\;g_{\nu\rho}),
\label{riemann2}
\eea
provided that the horizon is itself a constant curvature space
\bea
R_{ijkl}(h) = k(h_{ik}\;h_{jl}
- h_{il}\;h_{jk}).
\label{riemann3}
\eea
Thus, imposing the extra requirement that the $M=0$ solution
be a constant curvature spacetime, forces the horizon to be
a constant curvature space, and not simply Einstein.
However, once again there is no restriction on the sign of $k$.
Although the $M=0$ spacetime
is locally isometric to anti-de Sitter space, its topology
depends on the value of the $k$, and hence on the
topology  of the horizon. In particular, we have the three
possibilities of
elliptic horizons ($k=1$), flat horizons ($k=0$), and hyperbolic
horizons ($k=-1$). The hyperbolic $M=0$ solution appeared
in \cite{Mann2}.  We note from (\ref{f}) that the dominant behaviour
of the metric at infinity is determined by the cosmological constant term,
for any value of $M$. Since the $M=0$ solution is locally isometric
to anti-de Sitter space, we have a class of black hole solutions which
are asymptotically locally anti-de Sitter, for all values of $M$.

An instructive example is to consider
the situation in five dimensions. In this case, the horizon
is a three-dimensional compact manifold of constant
curvature, which may be either positive, negative,
or zero.
Now, it is known that any compact, constant curvature,
$3$-manifold $M^{3}$ can be expressed
as a quotient space, see \cite{Carlip} for example,
\bea
M^{3} = \tilde{M^{3}}/\Gamma,
\label{m3}
\eea
where the universal covering space $\tilde{M^{3}}$
is either the $3$-sphere (corresponding to $k=1$), the torus ($k=0$),
or hyperbolic $3$-space ($k=-1$), and $\Gamma$
is a discrete subgroup of the isometry group of
$\tilde{M}$. Thus, even in
the elliptic case, we have non-spherical possibilities;
for example, one may take the horizon to be a lens space.

To summarize, we have shown that the metric ansatz (\ref{ansatz})
and (\ref{f}) defines an Einstein spacetime
with negative cosmological
constant, and parameter $M$, provided that the horizon is itself
an Einstein space of positive, zero, or negative curvature.
The subclass of these solutions which are asymptotically locally
anti-de Sitter is obtained when the horizon is
a constant curvature space.

\section{Black Hole Thermodynamics}
In this section, we discuss the properties of the solutions
obtained, and in particular
their interpretation as black holes,
and the corresponding thermodynamics.

To implement the black hole interpretation, we
wish to restrict the parameters
so that the metric describes the exterior of a black
hole with a non-degenerate horizon.
This is achieved provided the polynomial $r^{d-3}l^{2}f$
has a simple positive
root $r_{+}$, such that $f(r) > 0$ for all $r > r_{+}$.
For $k=1$, and spherical topology, these solutions correspond
to those considered in \cite{Carter,HP}.
We simply note here that
the $k=1$ construction is not restricted to spherical
topology, but only require the horizon to be Einstein.
Thus, the thermodynamic analysis of \cite{HP} applies
to all the $k=1$ solutions.

For $k=0$, one can check that for $M>0$ there is always a simple positive
root of $r^{d-3}l^{2}f$ given by
\bea
r_{+} = (\omega_{d}Ml^{2})^{\frac{1}{d-1}}.
\label{r+}
\eea
In addition, $f(r) > 0$ for $r>r_{+}$. Thus, for $k=0$, we
have black hole solutions with toroidal topology.

The analysis for the case of $k=-1$ is more involved.
The four-dimensional case was studied in \cite{Mann}-\cite{Klemm},
where
it was shown that these solutions do indeed
have a black hole interpretation with the
horizon being a Riemann surface of
genus greater than one.
We content ourselves here by studying the five-dimensional case.
This also has particular relevance for the correspondence
with conformal field theory in four dimensions, which we shall
discuss.
In fact, in this case, one notes that we must only solve a quadratic
equation
\bea
(r^{2})^{2} -l^{2}r^{2} - \omega_{5}Ml^{2} = 0.
\label{quad}
\eea
It is convenient to define
\bea
r_{\mathrm{crit}} = \frac{l}{\sqrt{2}},\;\;
M_{\mathrm{crit}} = -\frac{l^{2}}{4\omega_{5}}.
\label{mcrit5d}
\eea
Now if $D = l^{4} + 4\omega_{5}Ml^{2} = 0$,
we see that $M = M_{\mathrm{crit}}$,
and there is a double zero, with no black hole
interpretation.
If $D<0$, which corresponds to $M < M_{\mathrm{crit}}$, there is no
real zero.
However, for $D>0$, i.e., $M > M_{\mathrm{crit}}$,
there are two possibilities. Firstly, for
$D>0$ and $M_{\mathrm{crit}} < M < 0$, there are two distinct
simple zeros. For $D>0$ and $M \geq 0$, there is one
simple zero. Hence, in these cases,
we have non-degenerate horizons with hyperbolic geometry.
The parameter $M$ is then determined by $f(r_{+}) = 0$, as
\bea
M = \frac{r_{+}^{2}}
{\omega_{5}}\left(-1 + \frac{r_{+}^{2}}{l^{2}}\right).
\label{mass1}
\eea
We see that $M_{\mathrm{crit}}$ corresponds to $M$
evaluated at $r_{\mathrm{crit}}$.
In addition, one notes that we can write
\bea
r^{2}l^{2}f = (r^{2} - r_{+}^{2})(r^{2} + r_{+}^{2} - l^{2}).
\eea
Since $r_{+} > r_{\mathrm{crit}}$, we see that $f$ is positive
for all $r > r_{+}$.
We remark that the seven-dimensional case also allows an explicit
analysis, since one must only solve a cubic equation in $r^{2}$.

Our aim now is to calculate the action of these black hole solutions,
and determine their thermodynamical properties. We may consider
any one of the above solutions which has an acceptable
horizon located at $r_{+}$.
The parameter $M$ is specified in terms of $r_{+}$ as
\bea
M = \frac{r_{+}^{d-3}}
{\omega_{d}}\left(k + \frac{r_{+}^{2}}{l^{2}}\right).
\label{mass}
\eea
The Euclidean Einstein action is proportional to the
spacetime volume, namely
\bea
I = -\frac{1}{16\pi G}\int d^{d}x\; \sqrt{g}(R - 2 \Lambda)
= \frac{(d-1)}{8 \pi G l^{2}}\int d^{d}x\;\sqrt{g},
\label{action1}
\eea
where the cosmological constant is
$\Lambda = -\frac{(d-1)(d-2)}{2l^{2}}$.
Since $I$ is infinite, we proceed in the standard way
and compare the action of the black hole with a convenient
background \cite{GH}. We note that the boundary terms which are
typically present in the action give zero contribution for
the cases under consideration.
Since we are working in the Euclidean formalism,
we must identify the imaginary time of the
solution with a period $\beta = 4 \pi/f^{\prime}(r_{+})$, namely
\bea
\beta = \frac{4 \pi l^{2}r_{+}}{(d-1)r_{+}^{2} + (d-3)kl^{2}}.
\label{beta}
\eea
It is worth highlighting some of the features
of these
black holes which depend on the value of $k$. For $k=1$,
it has been shown in \cite{HP} that the inverse temperature
$\beta$ given by (\ref{beta}) has a maximum value. Hence, the
black hole solutions only exist for temperatures greater than
a certain minimum value. The background geometry is
taken to be anti-de Sitter space with arbitrary Euclidean time
period.
However, for the case of $k=0$ and $k=-1$, it is easy to check from
(\ref{beta}) that there is no such minimum temperature; hence,
the $k=0$ and $k=-1$ black holes exist for all temperatures.
For $k=0$, the background can be taken to be the
$M=0$ solution \cite{Vanzo, Emparan}.
However, one does notice that for $k=-1$,
the requirement of positivity of temperature
enforces an inequality on the value of $r_{+}$,
namely that $r_{+} > r_{\mathrm{crit}}$, where
\bea
r_{\mathrm{crit}} = \left(\frac{d-3}{d-1}\right)^{\frac{1}{2}} l.
\label{rcrit}
\eea
Notice also that when $r_{+} = r_{\mathrm{crit}}$, we have
$M=M_{\mathrm{crit}}$, where
\bea
M_{\mathrm{crit}}
= - \left(\frac{2}{d-1}\right)
\left(\frac{d-3}{d-1}\right)^\frac{d-3}{2}\frac{l^{d-3}}{\omega_{d}}.
\label{mcrit}
\eea
Thus, the requirement of $r_{+} > r_{\mathrm{crit}}$ is
equivalent to the requirement that $M > M_{\mathrm{crit}}$, which
is needed in order to have a black hole interpretation.
Since the $M_{\mathrm{crit}}$ spacetime can be identified with
arbitrary Euclidean time period, we choose this as the background
geometry \cite{Vanzo, Emparan}.

The background spacetime is denoted by $X_{1}$ with
period $\beta_{0}$ and parameter $M_{\mathrm{crit}}$,
where $M_{\mathrm{crit}}$ is given by (\ref{mcrit}) for
$k=-1$, and is zero for $k=0,1$.
It should be stressed that since $f$ has a degenerate zero for
the $M_{\mathrm{crit}}$ spacetime with $k=-1$, the spacetime
has an internal infinity \cite{Brill2,Vanzo}.
As a result, the Killing horizons do
not have an interpretation as black hole horizons.
However, for the purposes of our analysis here,
we are restricting attention to the spacetime region
$r \geq r_{\mathrm{crit}}$, where $r_{\mathrm{crit}}$ is
given by (\ref{rcrit}). The asymptotic boundary of this
spacetime is then $S^{1} \times M^{d-2}$.
The black hole spacetime is denoted by $X_{2}$ with
period $\beta_{0}^{\prime}$, and an $r$-integration $r\geq r_{+}$.
To proceed further, we match $\beta_{0}$ and $\beta_{0}^{\prime}$
so that the geometry on the hypersurface at $r=R$ agrees.
This is achieved by taking
\bea
\beta_{0}\sqrt{k -\frac{\omega_{d}M_{\mathrm{crit}}}{r^{d-3}}
+ \frac{r^{2}}{l^{2}}} =
\beta_{0}^{\prime}\sqrt{k - \frac{\omega_{d}M}{r^{d-3}} +
\frac{r^{2}}{l^{2}}}.
\label{beta2}
\eea
Hence, the difference in the actions is
\bea
I \equiv I(X_{2}) - I(X_{1})
= \frac{\mathrm{Vol}(M^{d-2})}{16 \pi Gl^{2}}
\beta_{0}^{\prime} [- r_{+}^{d-1} + kl^{2} r_{+}^{d-3}]
- \beta_{0}^{\prime} M_{\mathrm{crit}}.
\label{action4}
\eea

Before analysing the phase structure of this action, let us
first note that the energy $E$ and entropy are given
in the usual way by
\bea
E = \frac{\partial I}{\partial \beta_{0}^{\prime}} =
M - M_{\mathrm{crit}}, \;\;
S = \beta_{0}^{\prime} E - I = \frac{\mathrm{Vol}(M^{d-2})}{4G}
r_{+}^{d-2}.
\eea

Finally, we note that the specific heat of the $k=0$ and
$k=-1$ black holes is positive. For $k=0$, we find
\bea
\frac{\partial E}{\partial T} = \frac{4 \pi}{\omega_{d}}r_{+}^{d-2},
\label{spheat1}
\eea
while for $k=-1$, we have
\bea
\frac{\partial E}{\partial T} =
\frac{4 \pi r_{+}^{d-2}}{\omega_{d}} \left[\frac{(d-1) r_{+}^{2}
-(d-3)l^{2}}{(d-1)r_{+}^{2} + (d-3)l^{2}}\right].
\label{spheat2}
\eea
In the latter case, we see that the specific heat is positive
provided $r_{+} > r_{\mathrm{crit}}$.
We contrast this with the case of $k=1$,
where for temperatures greater than the minimum
value there are two black holes, the smaller of which has negative
specific heat, the larger having positive specific heat.

\section{Correspondence with Conformal Field Theory}
There has been a recent flurry of activity
on the conjecture relating supergravity defined
on anti-de Sitter spaces and conformal field theory on
its boundary \cite{Malda}.
According to this conjecture in five dimensions, for example,
Type $IIB$ string theory defined on $adS_{5}\times S^{5}$ is equivalent
to the large $N$ limit of ${\cal N} = 4$ super Yang-Mills theory
with gauge group $SU(N)$ defined on $S^{4}$.
For our purposes here, we wish to concentrate on one
particular aspect, namely the relevance of the black hole solutions
to the phase structure of the conformal field theory.
Thus, we wish to study the conformal field theory at finite
temperature, defined on
a space $S^{1} \times M^{d-2}$.
In particular, it has been shown in \cite{Witten1,Witten2}
that one may use the adS/CFT correspondence to give a
holographic explanation of
the mass dependence of the Bekenstein-Hawking entropy
for the spherical black holes in anti-de Sitter space.
These correspond to the $k=1$ class with spherical horizon topology.
In \cite{Emparan}, the case of $M2$-branes wrapped around
higher genus Riemann surfaces was studied, leading to
information on the phase structure of the associated
conformal field theory
on $S^{1} \times \Sigma_{g}$, for $g\geq 1$.
Further aspects of the holography of this
correspondence have been studied in
\cite{Suss,Barbon}.

According to the prescription given in \cite{Witten1},
one computes the partition function of the boundary conformal field
theory on $S^{1} \times M^{d-2}$
by summing over contributions from Einstein
manifolds $M^{d}$ with negative cosmological constant which have
$S^{1} \times M^{d-2}$ as their boundary.
Thus, one has
\bea
Z_{CFT}(S^{1} \times M^{d-2}) = \sum_{i} e^{-I(X_{i})},
\eea
where $I$ is the supergravity action, and
in general one may need to consider the contribution
from several Einstein manifolds $X_{i}$ which have
$S^{1} \times M^{d-2}$ boundary.
It is then important to study the relative actions, and in this way
one obtains information on the existence of a phase transition.

For the case of conformal field theory on $S^{1}\times S^{d-2}$,
there are two such manifolds, namely
the black hole itself with compactified topology $B^{2}\times S^{d-2}$,
and thermal anti-de Sitter space corresponding to the background solution
with topology $S^{1}\times B^{d-1}$, where $B^{d}$ is a $d$-dimensional
ball with boundary $S^{d-1}$.
It was shown in \cite{Witten1,Witten2} that for large temperatures
the black hole dominated, while for small temperatures the thermal
anti-de Sitter space dominated
(due to the non-existence of the black
holes at  low temperatures). One may then compare the Bekenstein-Hawking
entropy with the entropy of the conformal field theory on
$S^{1}\times S^{d-2}$. The limit of high temperature,
$\beta_{0}^{\prime}\rightarrow 0$,
may be regarded as a high temperature system on $S^{d-2}$.
Conformal invariance then dictates that the entropy of this system
is of the order $(\beta_{0}^{\prime}/l)^{-(d-2)}$. However,
according to (\ref{beta}), one sees that for high temperature
we have $(\beta_{0}^{\prime}/l) \sim l/r_{+}$,
and hence the conformal field
theory entropy is of the order $(r_{+}/l)^{d-2}$, in agreement
with the Bekenstein-Hawking formula.
In this way, we obtain a microscopic understanding of the
mass dependence of the entropy for large black holes with $r_{+} >>l$.

We now wish to apply similar arguments to the topological black holes
considered here. For the $k=0$ case,
for which $M_{\mathrm{crit}} = 0$, the phase structure of the action
(\ref{action4}) is readily determined.
We have
\bea
I(X_{2}) - I(X_{1}) = \frac{\mathrm{Vol}(M^{d-2})}{16 \pi Gl^{2}}
\frac{4 \pi l^{2}}{(d-1)r_{+}}[- r_{+}^{d-1}].
\label{action5}
\eea
Now, for the $k=0$ case, we have seen that $r_{+} > 0$, and thus the
action is negative for all $r_{+}$.
From (\ref{beta}), we note
that low temperature corresponds to small $r_{+}$, while high temperature
corresponds to large $r_{+}$.
However, the action difference in (\ref{action5}) is negative
for all values of $r_{+}$, and hence we see no evidence of a phase
transition in this case. Thus, the black hole dominates
over its $M_{\mathrm{crit}}=0$ background counterpart
for all temperatures,
and we can once again
appeal to the correspondence with conformal field theory
to show that for high temperature (i.e., large $r_{+}$ black holes)
the entropy does indeed scale holographically
in the correct manner as  $r_{+}^{d-2}$.
Note also that we have seen that the toroidal black holes
have positive
specific heat, which is consistent with the conformal
field theory correspondence.

In order to determine the phase structure for the $k=-1$ case,
we should first recall that the $M_{\mathrm{crit}}$ background spacetime
has an internal infinity. In the evaluation of the action,
we restricted attention to the region
$r \geq r_{\mathrm{crit}}$ for the background and $r \geq r_{+}$
for the black hole. Thus, the partition function  of the
corresponding conformal field theory
is sensitive only to these regions. We shall adopt this as
an assumption on how to treat these internal infinities.
The action difference is
\bea
I(X_{2}) - I(X_{1}) = \frac{\mathrm{Vol}(M^{d-2})}{16 \pi Gl^{2}}
\beta_{0}^{\prime}[- r_{+}^{d-1} - l^{2} r_{+}^{d-3}]
-\beta_{0}^{\prime}M_{\mathrm{crit}}.
\eea
We have seen that for $r_{+} > r_{\mathrm{crit}}$, the temperature
is positive. Thus, for all $r_{+} > r_{\mathrm{crit}}$, the action
is negative, tending to zero as $r_{+} \rightarrow r_{\mathrm{crit}}$.
As a result, the black hole again dominates
over the $M_{\mathrm{crit}}$ background for all
temperatures, with no phase transition
occurring.
Once again, we see from (\ref{beta}) that low temperature corresponds
to small $r_{+}$, while high temperature corresponds to
large $r_{+}$. Thus, using the conformal field theory correspondence,
we again have a holographic explanation of the Bekenstein-Hawking
entropy formula for large black holes. We stress that these
conclusions rely on our assumption of how to treat the internal
infinity of the background spacetime.

\section{Conclusions}
As indicated in \cite{Witten1}, the conformal
theory defined on $S^{1}\times M^{d-2}$ takes two forms.
For spinors which are periodic on $S^{1}$,
we have
\bea
Z_{1}(S^{1} \times M^{d-2}) = \mathrm{Tr}\; (-1)^{F} e^{-\beta H},
\eea
while anti-periodic spinors yield
\bea
Z_{2}(S^{1}\times M^{d-2}) = \mathrm{Tr}\;e^{-\beta H},
\eea
where $H$ is the Hamiltonian.
For the case of black holes with spherical
topology studied in \cite{Witten1,Witten2},
the thermal anti-de Sitter space background
has topology $S^{1} \times B^{d-1}$, and being non-simply connected
contributes  to both $Z_{1}$ and $Z_{2}$.
However, the black holes are simply connected having topology
$B^{2} \times S^{d-2}$; thus, they contribute only to $Z_{2}$
where the phase transition was observed.
For the black hole solutions constructed here, we see that
they have compactified topology $B^{2} \times M^{d-2}$,
where $M^{d-2}$ is in general non-simply connected.
Thus, they will contribute to both $Z_{1}$ and $Z_{2}$, and it would
be interesting to check their role in understanding the phase
structure of the former.

Finally, we remark that for the hyperbolic case,
the numerical coefficient in the Bekenstein-Hawking
entropy is the hyperbolic volume, which
is  known to be a topological invariant in three-manifold theory,
see for example \cite{Carlip,Carlip2}, and also \cite{Carlip3} for
a discussion in four dimensions.
Since the correspondence with conformal field theory
provided sufficient information
to fix the mass dependence of the entropy, it would be interesting
to see if the topological nature of the horizon volume can be
used to fix the numerical coefficient.

\noindent {\large \bf Acknowledgements}\\
I would like to thank R. Emparan for valuable correspondence
on a previous version of the manuscript, the Theory Division
at CERN for hospitality, and Forbairt for support under grant number
SC/96/603.


\begin{thebibliography}{99}
\bibitem{Malda} J. Maldacena, {\em The Large $N$ Limit of
Superconformal Field Theories and Supergravity},
Adv. Theor. Math. Phys. 2 (1998) 231, hep-th/9711200.
\bibitem{Gub} S.S. Gubser, I.R. Klebanov and A.W. Peet,
{\em Entropy and Temperature of Black $3$-Branes},
Phys. Rev. D54 (1996) 3915; hep-th/9602135.
\bibitem{Kleb} I.R. Klebanov, {\em World Volume Approach to
Absorption by Non-dilatonic Branes}, Nucl. Phys. B496 (1997) 231;
hep-th/9702076.
\bibitem{Tseytlin} S.S. Gubser, I.R. Klebanov and A.A. Tseytlin,
{\em String Theory and Classical Absorption by Three-Branes},
Nucl. Phys. B499 (1997) 217; hep-th/9703040.
\bibitem{Gub2} S.S. Gubser and I.R. Klebanov, {\em Absorption by Branes
and Schwinger Terms in the World Volume Theory},
Phys. Lett. B413 (1997) 41; hep-th/9708005.
\bibitem{Strom} J. Maldacena and A. Strominger, {\em Semiclassical
Decay of Near Extremal Fivebranes}, J. High Energy Phys.
12 (1997) 008, hep-th/9710014.
\bibitem{Gibbons} G.W. Gibbons and P.K. Townsend,
{\em Vacuum Interpolation in Supergravity via Super $p$-branes},
Phys. Rev. Lett. 71 (1993) 3754, hep-th/9307049.
\bibitem{Duff} M.J. Duff, G.W. Gibbons and P.K. Townsend, {\em
Macroscopic Superstrings as Interpolating Solitons}, Phys. Lett.
B332 (1994) 321; hep-th/9405124.
\bibitem{Poly} S.S. Gubser, I.R. Klebanov and A.M. Polyakov,
{\em Gauge Theory Correlators from Non-critical String Theory},
Phys. Lett. B. 428 (1998) 105, hep-th/9802109.
\bibitem{Witten1} E. Witten, {\em Anti-de Sitter Space and Holography},
Adv. Theor. Math. Phys. 2 (1998) 253, hep-th/9802150.
\bibitem{Witten2} E. Witten, {\em Anti-de Sitter Space, Thermal
Phase Transition, and Confinement in Gauge Theories},
Adv. Theor. Math. Phys. 2 (1998) 505, hep-th/9803131.
\bibitem{Carter} B. Carter, Commun. Math. Phys. 10 (1968) 280.
\bibitem{HP} S.W. Hawking and D.N. Page, {\em Thermodynamics
of Black Holes in Anti-de Sitter Spaces}, Commun. Math. Phys.
87 (1983) 577; J.D. Brown, J. Creighton and R.B. Mann, {\em Temperature,
Energy, and Heat Capacity of Asymptotically Anti-de Sitter Black Holes},
Phys. Rev. D50 (1994) 6394; gr-qc/9405007.
\bibitem{Mann} R.B. Mann, {\em Pair Production of Topological
Anti-de Sitter Black Holes},
Class. Quantum Grav. 14 (1997) L109; gr-qc/9607071.
\bibitem{Brill2} D.R. Brill, J. Louko and P. Peld\'{a}n,
{\em Thermodynamics of $(3+1)$-Dimensional Black Holes with Toroidal
or Higher Genus Horizons}, Phys. Rev. D56 (1997) 3600, gr-qc/9705012.
\bibitem{Vanzo} L. Vanzo, {\em Black Holes with Unusual Topology},
Phys. Rev. D56 (1997) 6475, gr-qc/9705004.
\bibitem{Beng} S. {\AA}minneborg, I. Bengtsson,
S. Holst and P. Peld\'{a}n, {\em Making Anti-de Sitter Black Holes},
Class. Quantum Grav. 13 (1996) 2707; gr-qc/9604005.
\bibitem{Brill1} D.R. Brill, {\em Multi-Black-Holes in $3D$
and $4D$ Anti-de Sitter Spacetimes}, Helv. Phys. Acta 69 (1996)
249; gr-qc/9608010.
\bibitem{Klemm} D. Klemm, V. Moretti and L. Vanzo, {\em Rotating
Topological Black Holes}, Phys. Rev. D57 (1998) 6127; gr-qc/9710123.
\bibitem{Lemos1}
J.P.S. Lemos, {\em Three-Dimensional Black Holes and Cylindrical
General Relativity}, Phys. Lett. B. 353 (1995) 46, gr-qc/9404041;
J.P.S. Lemos, {\em Two-Dimensional Black Holes
and Planar General Relativity}, Class. Quantum Grav. 12 (1995) 1081,
gr-qc/9407024;
J.P.S. Lemos and V.T. Zanchin, {\em Rotating
Charged Black Strings in General Relativity},
Phys. Rev. D54 (1996) 3840, hep-th/9511188.
\bibitem{Huang} C.-G. Huang and C.-B. Liang, {\em A Torus-Like
Black Hole}, Phys. Lett. A201 (1995) 27.
\bibitem{Cai} R.-G. Cai and Y.-Z. Zhang, {\em Black Plane
Solutions in Four-Dimensional Spacetimes}, Phys. Rev. D54
(1996) 4891; gr-qc/9609065.
\bibitem{Mann2} R.B. Mann, {\em Topological Black Holes - Outside
Looking In}, in {\em Internal Structure of Black Holes and Spacetime
Singularities}, ed. L. Burko and A. Ori, Technion University Press,
1998; gr-qc/9709039.
\bibitem{Ban} M. Ba\~{n}ados, {\em Constant Curvature Black Holes},
Phys. Rev. D57 (1998) 1068, gr-qc/9703040;
M. Ba\~{n}ados, A. Gomberoff and C. Mart\'{\i}nez,
{\em Anti-de Sitter Space and Black Holes}, hep-th/9805087.
\bibitem{HE} S.W. Hawking and G.F.R. Ellis, {\em The Large Scale
Structure of Space-time}, Cambridge University Press, Cambridge, 1973.
\bibitem{Carlip3} S. Carlip, {\em Dominant Topologies in Euclidean
Quantum Gravity}, Class. Quantum Grav. 15 (1998) 2629, gr-qc/9710114.
\bibitem{Carlip} S. Carlip, {\em The Sum over Topologies in
Three-Dimensional
Euclidean Quantum Gravity}, Class. Quantum Grav. 10 (1993) 207,
hep-th/9206103.
\bibitem{GH} G.W. Gibbons and S.W. Hawking, {\em Action Integrals and
Partition Functions in Quantum Gravity}, Phys. Rev. D15 (1977) 2752.
\bibitem{Emparan} R. Emparan, {\em $AdS$ Membranes Wrapped
on Surfaces of Arbitrary Genus}, Phys. Lett. B432 (1998) 74,
hep-th/9804031.
\bibitem{Suss} L. Susskind and E. Witten, {\em The Holographic Bound
in Anti-de Sitter Space}, hep-th/9805114.
\bibitem{Barbon} J.L.F. Barb\'{o}n and E. Rabinovici, {\em Extensivity
Versus Holography in Anti-de Sitter Spaces}, hep-th/9805143.
\bibitem{Carlip2} S. Carlip, {\em Entropy Versus Action in the
$(2+1)$-Dimensional Hartle-Hawking Wave Function}, Phys. Rev. D46
(1992) 4397; hep-th/9205022.
\end{thebibliography}
\end{document}